\documentclass[12pt,a4paper]{article}
\usepackage{cmap}
\usepackage[T2A]{fontenc}
\usepackage[cp1251]{inputenc}
\usepackage[english, russian]{babel}
\usepackage{amsfonts,amsmath,amssymb,amsthm,longtable,hhline}
\usepackage{bm}
\usepackage{cite}
\usepackage{soul}
\usepackage{color, xcolor}
\usepackage{multicol}
\usepackage[colorlinks,linkcolor=red,citecolor=green]{hyperref}
\usepackage{tabularx,caption,multirow}

\makeatletter
\renewcommand{\@biblabel}[1]{#1.} 
\makeatother
 \newcounter{remark}
 \newtheorem{Remark}[remark]{Remark}

\let\OLDthebibliography\thebibliography
\renewcommand\thebibliography[1]{
  \OLDthebibliography{#1}
  \setlength{\parskip}{0pt}
  \setlength{\itemsep}{5pt}
}

\let\ds=\displaystyle
\let\ts=\textstyle
\let\bl=\bigl \let\br=\bigr
\let\Bl=\Bigl \let\Br=\Bigr
 
\def\arb{is an arbitrary constant}
\def\arbs{are arbitrary constants}
\def\arbf{is an arbitrary function}
\def\arbfs{are arbitrary functions}
\def\fracskip{\mskip 1mu \relax}
\def\nfrac#1#2{{\fracskip#1\fracskip\over\fracskip#2\fracskip}}
\def\dfrac#1#2{{\ds\nfrac{#1}{#2}}}
\def\tfrac#1#2{{\ts\nfrac{#1}{#2}}}
\let\frac=\nfrac
\def\pd#1#2{\dfrac{\partial#1}{\partial#2}}
\def\pdd#1#2#3{\ifx#2#3\pd{^2#1}{#2^2}\else\pd{^2#1}{#2\partial#3}\fi }

\newcommand{\clh}[1]{\colorbox{yellow}{#1}}%
\newcommand{\clhp}[1]{\colorbox{yellow}{\parbox{\textwidth}{#1}}}%
\makeatletter
\g@addto@macro\bfseries{\boldmath}
\makeatother

\begin{document}
\large 

\centerline{\bf Exact Solutions and Reductions of}
\centerline{\bf Nonlinear Diffusion PDEs of Pantograph Type\clh{$^*$}}

\bigskip

\centerline{Andrei D. Polyanin, Vsevolod G. Sorokin}
\bigskip
\centerline{\it Ishlinsky Institute for Problems in Mechanics RAS,}
\centerline{\it 101 Vernadsky Avenue, bldg~1, 119526 Moscow, Russia}
\bigskip
\bigskip

\let\thefootnote\relax\footnotetext{
\hskip-20pt\clhp{$^*$ This is a preprint of the article
A.D.\! Polyanin, V.G.\! Sorokin, Nonlinear pantograph-type diffusion PDEs: Exact solutions and the principle of analogy, \textit{Mathematics}, 2021, 9(5), 511; doi:10.3390/math9050511.
}}

\textbf{Abstract.} We study nonlinear pantograph-type reaction-diffusion PDEs, which, in addition to the unknown
$u=u(x,t)$, also contain the same functions with dilated or contracted arguments of the form
$w=u(px,t)$, $w=u(x,qt)$, and $w=u(px,qt)$,
where $p$ and $q$ are the free scaling parameters (for equations with proportional delay we have $0<p<1$, $0<q<1$).
A brief review of publications on pantograph-type ODEs and PDEs and their applications is given.
Exact solutions and reductions of various types of such nonlinear partial functional differential equations are described for the first time. We present examples of nonlinear
pantograph-type PDEs with proportional delay, which admit traveling-wave and self-similar solutions (note that PDEs with constant delay do not have self-similar solutions).
Additive, multiplicative and functional separable solutions, as well as some other exact solutions are also obtained.
Special attention is paid to nonlinear pantograph-type PDEs of a rather general form, which contain one or two arbitrary functions.
In total, more than forty nonlinear pantograph-type reaction-diffusion PDEs with dilated or contracted arguments, admitting exact solutions,
have been considered.
Multi-pantograph nonlinear PDEs are also discussed.
The principle of analogy is formulated, which makes it possible to efficiently construct exact solutions of nonlinear pantograph-type PDEs.
A number of exact solutions of more complex nonlinear functional differential equations with varying delay, which arbitrarily depends on time or spatial coordinate,
are also described.
The presented equations and their exact solutions can be used to formulate test problems designed to evaluate the accuracy
of numerical and approximate analytical methods for solving the corresponding nonlinear initial-boundary value problems
for PDEs with varying delay. The principle of analogy  allows finding solutions to other nonlinear pantograph-type PDEs
(including nonlinear wave-type PDEs and higher-order equations).

\bigskip


\section*{Contents}

\hfill\begin{minipage}{\dimexpr\textwidth-\parindent}

\begin{small}

\hyperlink{hlink-s:1}{\textbf{1. Introduction}}

\hyperlink{hlink-ss:1.1}{\hspace{1.25em}1.1. Differential Equations with Constant Delay}

\hyperlink{hlink-ss:1.2}{\hspace{1.25em}1.2. Pantograph-Type ODEs and PDEs and Their Applications}

\hyperlink{hlink-ss:1.3}{\hspace{1.25em}1.3. Concept of `Exact Solution' for Nonlinear Pantograph-Type PDEs}

\medbreak
\hyperlink{hlink-s:2}{\textbf{2. Solutions of Pantograph-Type PDEs with $u=u(x,t)$ and $w=u(px,qt)$}}

\hyperlink{hlink-ss:2.1}{\hspace{1.25em}2.1. Equations Containing Free Parameters}

\hyperlink{hlink-ss:2.2}{\hspace{1.25em}2.2. Equations Linear in Derivatives and Containing One Arbitrary Function}

\hyperlink{hlink-ss:2.3}{\hspace{1.25em}2.3. More Complex Nonlinear Equations Containing One Arbitrary Function}

\hyperlink{hlink-ss:2.4}{\hspace{1.25em}2.4. Nonlinear Equations Containing Two Arbitrary Functions}

\medbreak
\hyperlink{hlink-s:3}{\textbf{3. Solutions of Pantograph-Type PDEs with $u=u(x,t)$ and $w=u(x,qt)$}}

\hyperlink{hlink-ss:3.1}{\hspace{1.25em}3.1. Equations Containing Free Parameters}

\hyperlink{hlink-ss:3.2}{\hspace{1.25em}3.2. Equations Linear in Derivatives and Containing One Arbitrary Function}

\hyperlink{hlink-ss:3.3}{\hspace{1.25em}3.3. More Complex Nonlinear Equations Containing One Arbitrary Function}

\hyperlink{hlink-ss:3.4}{\hspace{1.25em}3.4. Nonlinear Equations Containing Two Arbitrary Functions}

\medbreak
\hyperlink{hlink-s:4}{\textbf{4. Solutions of Pantograph-Type PDEs with $u=u(x,t)$ and $w=u(px,t)$}}

\hyperlink{hlink-ss:4.1}{\hspace{1.25em}4.1. Equations Linear in Derivatives}

\hyperlink{hlink-ss:4.2}{\hspace{1.25em}4.2. More Complex Nonlinear Equations}

\medbreak
\hyperlink{hlink-s:5}{\textbf{5. Some Generalizations}}

\hyperlink{hlink-ss:5.1}{\hspace{1.25em}5.1. Nonlinear Multi-Pantograph Type PDEs}

\hyperlink{hlink-ss:5.2}{\hspace{1.25em}5.2. Nonlinear Pantograph-Type PDEs with Dilation/Contraction of Arguments in the Derivative}

\hyperlink{hlink-ss:5.3}{\hspace{1.25em}5.3. Principle of Analogy of Solutions}

\medbreak
\hyperlink{hlink-s:6}{\textbf{6. Brief Conclusions}}

\medbreak
\hyperlink{hlink-s:7}{\textbf{References}}

\end{small}

\xdef\tpd{\the\prevdepth}
\end{minipage}

\hypertarget{hlink-s:1}{}
\section{Introduction}\label{s:1}

\hypertarget{hlink-ss:1.1}{}
\subsection{Differential Equations with Constant Delay}

In natural science, an important role is played by the study of the hereditary properties of nonlinear systems
of different nature, when the rate of change of the  unknown depends not only on the state at a given time,
but also on the previous evolution of the process.
In a particular case, the state of the system is determined not by its entire history, but by the current moment and some moment in the past.
Such systems are called delay systems. In the simplest case, the differential equations modelling such systems,
in addition to the unknown function $u(t)$, also contain the function $u(t-\tau)$, where $t$ is the time and $\tau>0$ is the delay \cite{ofe1,els,mys1972}.
The delay is generally assumed constant.

When modeling nonlinear systems with constant delay and two independent variables $x$ and $t$, where $x$ is the spatial coordinate,
reaction-diffusion equations of the following form are most often encountered \cite{wu1996}:
 \begin{align}
 u_t=au_{xx}+F(u,w),\quad \ w=u(x,t-\tau).
 \label{eq:000}
\end{align}
The presence of delay significantly complicates the analysis of equations of the form \eqref{eq:000}. Such equations admit traveling-wave solutions $u=u(z)$, where $z=x+\lambda t$
(e.g, see \cite{wu1996,mei2009,lv2015,polsor2015aml}), but do not admit self-similar solutions $u=t^\beta \varphi(x t^\lambda)\,$
(recall that many PDEs without delay have self-similar solutions).


More complex than traveling-wave, exact solutions of nonlinear delay reaction-diffusion equations are obtained in
\cite{mel2008,pol2013,pol2014a,pol2014c,pol2014b,pol2014d,pol2014ee,pol2014f,pol2015bb,pol2019a,polsor2021,polsor2020**}.
Exact solutions of nonlinear delay equations of the Klein---Gordon type and other nonlinear hyperbolic equations are given in
\cite{pol2014g,polsorvyaz2015,long2016,long2017,polsor2020a,polsor2021,polsor2020**}.
Some exact solutions of differential-difference
equations of a viscous incompressible fluid (which generalize the Navier---Stokes equations) are described in \cite{pol2014h}.
Numerical methods for solving delay PDEs are discussed, for example, in \cite{sha2009,rih2010,sch2020}.

It is important to note that delay differential equations have a number of specific qualitative features \cite{wu1996,jor2008,polsorvyaz2015}
that are not inherent in equations without delay.

\hypertarget{hlink-ss:1.2}{}
\subsection{Pantograph-Type ODEs and PDEs and Their Applications}

In some cases, the delay may be time dependent, i.e. $\tau=\tau(t)$, where $\tau(t)>0$ is a given function~\cite{els}.

To illustrate this, consider a first-order linear ODE with varying delay proportional to the independent variable:
\begin{align}
u'_t=au+bw,\quad w=u(pt),
\label{ode1:1**}
\end{align}
which for $p>0$ ($p\not=1$) is called \textit{the pantograph equation}.

For $0<p<1$, the equation describes the dynamics of
a current collection system for an electric locomotive \cite{ock1971} and is a special case of an ODE with varying delay for
$\tau(t)=(1-p)t$.
The function $u(pt)$ included in the pantograph equation \eqref{ode1:1**} differs from the function $u(t)$ by dilation along the $t$-axis by $1/p$ times.

The pantograph equation and more complex related ODEs and PDEs that contain the unknown functions with dilation (for $0<p<1$) or contraction (for $p>1$)
of arguments are used in mathematical modeling of various processes in biology \cite{hal1989,hal1991,der2012,zai2015,efe2018},
astrophysics \cite{amb1944}, electrodynamics  \cite{deh2008},
population theory \cite{aje1992}, number theory \cite{mah1940}, stochastic games \cite{fer1972}, graph theory \cite{rob1973}, risk and queue theory \cite{gav1964}, and
theory of neural networks  \cite{zha2013}. Note that in \cite{hal1989,amb1944,zai2015,efe2018} the equations are derived for the case $p>1$.

Analysis of pantograph-type ODEs and construction of approximate analytical solutions are carried out, for example,
in \cite{fox1971,ise1993,kat1971,liu2004,ock1971,vanbru2011, yus2013,reu2015,isi2016,pat2017,bah2020,hou2020, alra2020}.
Numerical methods for solving such equations are discussed in
\cite{liu1996,bellen1997, koto1999,bel2002,gug2003,xu2004, eva2005,li2005,liu2006,sez2007,sez2008,saad2009, bru2010,shak2010,yus2010,gul2011,yal2011,yuz2011,sed2012,toh2013, doh2014,li2014,wan2015,wang2017,yang2018, yang2021}.
Thus, in general, approximate analytical and numerical methods for solving pantograph-type ODEs can currently be considered fairly well developed.

Significantly fewer publications are devoted to the analysis and solving of pantograph-type PDEs.
In \cite{zai2015}, a linear first-order pantograph-type PDE is used to model the growth and division of cells structured by size.
Its solution is sought in the form of a series, the terms of which are determined by solving simpler equations without dilated arguments.
In \cite{efe2018}, a more complex linear pantograph-type reaction-diffusion equation is investigated
(obtained by adding a diffusion term to the equation that was analyzed in \cite{zai2015}).
In \cite{liu2018}, solutions of linear pantograph-type heat and wave equations are obtained using the method of separation of variables.
In \cite{ros2006,ros2017,sku2016}, the problems of unique solvability and smoothness of linear
boundary value problems for elliptic PDEs
with dilation or contraction of arguments in higher derivatives are considered.
Analytical methods for solving some linear and nonlinear PDEs with proportional
delays are discussed in \cite{aba2011,gro2020,aks2021}.
In \cite{sol2015}, a finite-difference scheme for the numerical integration of first-order PDEs with constant delay in $t$ and proportional delay in $x$ is constructed.
Papers \cite{suk2016,bek2020,tang2021} are devoted to numerical methods for solving pantograph-type PDEs with proportional delay \cite{suk2016,bek2020}
and more complex varying delay \cite{tang2021}.


This article describes different classes of exact solutions to
nonlinear pantograph-type reaction-diffusion equations of the form
\begin{align}
u_{t}=[g(u)u_x]_x+F(u,w),\label{eq:007}
\end{align}
where $u=u(x,t)$ and $w=u(px,qt)$, $p>0$ and $q>0$ ($p$ and $q$ cannot be equal to $1$ at the same time).

\begin{Remark}
Exact solutions of `ordinary' nonlinear diffusion type equations (without dilated or contracted arguments) can be found, for example, in
\cite{pol2012,dor1982,nuc1992,kudr1993,gal1994,ibr1994,doy1998,puc2000,est2002,kap2003,gal2007,van2007,van2012,bro2016,cher2018,bra2019,goa2019,kos2019,pol2019d,pol2019aa,
kos2020,opa2020,pol2019ab,pol2020aa,pol2020ab}.
\end{Remark}

In most cases, exact solutions of pantograph-type PDEs of the form \eqref{eq:007} are obtained as follows:
a solution to a pantograph-type equation is sought in the same form as a solution to a simpler equation without delay,
which is obtained from \eqref{eq:007} for $w=u$ (for details, see Section~\ref{ss:5.3}).
More complex exact solutions are found using various modifications of the functional constraints method \cite{pol2014a, pol2014f}.
In addition, to construct some exact solutions of pantograph-type equations, we use exact solutions of nonlinear PDEs with varying delay of general form,
which were given in \cite{pol2013,pol2014c,pol2014f}.

As a result of the analysis, we have managed to find many nonlinear pantograph-type reaction-diffusion equations,
which admit exact traveling-wave solutions, self-similar solutions, additive, multiplicative, generalized, and functional separable solutions,
as well as some other exact solutions.
Special attention is paid to nonlinear pantograph-type PDEs
of a rather general form that contain arbitrary functions (it is these equations and their solutions that are
of greatest interest for testing numerical and approximate analytical
methods for solving the corresponding nonlinear
initial-boundary value problems for PDEs with varying delay).

\hypertarget{hlink-ss:1.3}{}
\subsection{Concept of `Exact Solution' for Nonlinear Pantograph-Type PDEs}

In this article, the term `exact solution' for nonlinear  pantograph-type PDEs (with dilation or contraction of one or several independent variables) and equations with varying delay will be used in cases where the solution is expressed:

(i) in terms of elementary functions, functions included in the equation (this is necessary when the equation contains arbitrary functions), and indefinite and/or definite integrals;

(ii) in terms of solutions of ODEs without delay or systems of such equations;

(iii) in terms of solutions of pantograph-type ODEs and ODEs with constant or varying delay or systems of such equations.

Combinations of solutions from paragraphs (i)--(iii) are allowed. In case (i), an exact solution can be represented in explicit, implicit or parametric form.
This definition generalizes the concept of `exact solution' that is used for nonlinear PDEs with constant delay in \cite{pol2014a,pol2014b}  and for PDEs without delay in \cite{pol2012}.

In what follows, unless otherwise specified, in nonlinear pantograph-type PDEs, it is assumed that $p$ and $q$ are the free scaling parameters
($p$ and $q$ are positive and cannot be equal to $1$ at the same time), $f(z)$ and $g(z)$ \arbfs, and $a$, $b$, $c$, etc. are arbitrary real constants.
The unknown function is denoted by $u$, and the unknown function with dilation or contraction of one or more arguments is denoted by $w$.

\hypertarget{hlink-s:2}{}
\section{Solutions of Pantograph-Type PDEs That Contain Unknown Functions $u=u(x,t)$ and $w=u(px,qt)$}\label{s:2}

\hypertarget{hlink-ss:2.1}{}
\subsection{Equations Containing Free Parameters}\label{ss:2.1}

\textit{\textbf{Equation 1.}} The equation with power-law nonlinearity
\begin{align}
u_t=au_{xx}+bw^k,\quad \ w=u(px,qt),
\label{eq:001}
\end{align}
for $k\not=1$ admits the self-similar solution
\begin{align}
u(x,t)=t^{\ts\frac 1{1-k}}U(z), \quad \ z=xt^{-1/2},
\label{eq:002}
\end{align}
where the function $U=U(z)$ is described by the nonlinear pantograph-type ODE:
\begin{align}
aU''_{zz}+\tfrac12zU'_z-\frac 1{1-k}U+bq^{\frac k{1-k}}W^k=0,\quad \ W=U(sz),\quad \ s=pq^{-1/2}.
\label{eq:003}
\end{align}

\begin{Remark}
It is interesting to note that the ODE with proportional delay \eqref{eq:001} for\break
$0<p,q<1$ in the special case $p=q^{1/2}$ has an exact solution
expressed in terms of the solution of the ODE without delay  \eqref{eq:003} with $s=1$;
for $p<q^{1/2}$, Eq.~\eqref{eq:001} reduces to the delay ODE with $s<1$;
and for $p>q^{1/2}$, to the ODE with contracted argument for $s>1$. Moreover, a solution of the ODE~\eqref{eq:001}
for $p,q>1$ for appropriate values of the parameters $p$ and $q$ can also be expressed in terms of the solution of the ODE with delay ($s<1$),
without delay ($s=1$), and with contracted argument ($s>1$).

The equations and their solutions given below in Section~\ref{s:2} have similar qualitative features.
\end{Remark}

\medskip
\textit{\textbf{Equation 2.}} The more complex equation with power-law nonlinearity
\begin{align*}
u_t=au_{xx}+bu^mw^k,\quad \ w=u(px,qt),
\end{align*}
for $k\not=1-m$ admits the self-similar solution
\begin{align*}
u(x,t)=t^{\ts\frac 1{1-m-k}}U(z), \quad \ z=xt^{-1/2},
\end{align*}
where the function $U=U(z)$ is described by the nonlinear pantograph-type ODE:
\begin{align*}
&aU''_{zz}+\tfrac12zU'_z-\frac 1{1-m-k}U+bq^{\frac k{1-m-k}}U^mW^k=0,\\
&W=U(sz),\quad \ s=pq^{-1/2}.
\end{align*}

\medskip
\textit{\textbf{Equation 3.}}  The equation with exponential nonlinearity
\begin{align*}
u_t=au_{xx}+be^{\lambda w},\quad \ w=u(px,qt),
\end{align*}
admits the exact solution
\begin{align*}
u(x,t)=U(z)-\frac 1\lambda \ln t,\quad \ z=xt^{-1/2},
\end{align*}
where the function $U=U(z)$ is described by the nonlinear pantograph-type ODE:
\begin{align*}
&aU''_{zz}+\frac 12zU'_z+\frac 1\lambda+\frac bq e^{\lambda W}=0,\\
&W=U(sz),\quad \ s=pq^{-1/2}.
\end{align*}

\medskip
\textit{\textbf{Equation 4.}}  The more complex equation with exponential nonlinearity
\begin{align*}
u_t=au_{xx}+be^{\mu u+\lambda w},\quad \ w=u(px,qt),
\end{align*}
for $\mu\not=-\lambda$ admits the exact solution
\begin{align*}
u(x,t)=U(z)-\frac 1{\mu+\lambda} \ln t,\quad \ z=xt^{-1/2},
\end{align*}
where the function $U=U(z)$ is described by the nonlinear pantograph-type ODE:
\begin{align*}
&aU''_{zz}+\frac 12zU'_z+\frac 1{\mu+\lambda}+bq^{\ts-\frac{\lambda}{\mu+\lambda}} e^{\mu U+\lambda W}=0,\\
&W=U(sz),\quad \ s=pq^{-1/2}.
\end{align*}

\medskip
\textit{\textbf{Equation 5.}}  The equation with logarithmic nonlinearity
\begin{align}
u_t=au_{xx}+u(b\ln u+c\ln w+d),\quad \ w=u(px,qt),
\label{eq:2000}
\end{align}
admits the multiplicative separable solution
\begin{align}
u(x,t)=\varphi(x)\psi(t),
\label{eq:2000**}
\end{align}
where the functions $\varphi=\varphi(x)$ and $\psi=\psi(t)$ are described by the nonlinear pantograph-type ODEs:
\begin{align*}
&a\varphi''_{xx}+\varphi(b\ln \varphi+c\ln\bar\varphi+d)=0,\quad \ \bar\varphi=\varphi(px);\\
&\psi'_t=\psi(b\ln \psi+c\ln\bar\psi),\quad \ \bar\psi=\psi(qt).
\end{align*}

\begin{Remark}
Eq. \eqref{eq:2000} and its solution can be substantially generalized if the pantograph term
 $w=u(px,qt)$ is replaced by $w=u(\xi(x),\eta(t))$, where $\xi(x)$ and $\eta(t)$ \arbfs \ (the solution is sought in the same form \eqref{eq:2000**}).
\end{Remark}

\hypertarget{hlink-ss:2.2}{}
\subsection{Equations Linear in Derivatives and Containing One Arbitrary Function}\label{ss:2.2}

\textit{\textbf{Equation 6.}}  The equation containing an arbitrary function
\begin{align*}
u_{t}=au_{xx}+u^nf(w/u),\quad \ w=u(px,qt),
\end{align*}
for $n\not=1$ admits the self-similar solution
\begin{align*}
u(x,t)=t^{\ts\frac 1{1-n}}U(z),\quad \ z=xt^{-1/2},
\end{align*}
where the function $U=U(z)$ satisfies the second-order pantograph-type ODE:
\begin{align*}
&\frac 1{1-n}U-\frac 12zU'_z=aU''_{zz}+U^nf(q^{\frac{1}{1-n}}W/U),\\
&W=U(sz),\quad \ s=pq^{-1/2}.
\end{align*}

\medskip
\textit{\textbf{Equation 7.}}  The equation containing an arbitrary function
\begin{align*}
u_t=au_{xx}+e^{\lambda u}f(u-w),\quad \ w=u(px,qt),
\end{align*}
admits the exact solution
\begin{align*}
u(x,t)=U(z)-\frac 1\lambda \ln t,\quad \ z=xt^{-1/2},
\end{align*}
where the function $U=U(z)$ is described by the nonlinear pantograph-type ODE:
\begin{align*}
&aU''_{zz}+\frac 12zU'_z+\frac 1\lambda+e^{\lambda U}f\Bl(U-W+\frac 1\lambda\ln q\Br)=0,\\
&W=U(sz),\quad \ s=pq^{-1/2}.
\end{align*}

\hypertarget{hlink-ss:2.3}{}
\subsection{More Complex Nonlinear Equations Containing One Arbitrary Function}\label{ss:2.3}

\textit{\textbf{Equation 8.}}  The equation with varying transfer coefficient of power-law type
\begin{align}
u_{t}=a(u^ku_x)_x+u^nf(w/u),\quad \ w=u(px,qt),
\label{eq-z}
\end{align}
for $n\not=1$ admits the self-similar solution
\begin{align*}
u(x,t)=t^{\ts\frac 1{1-n}}U(z),\quad \ z=xt^{\ts\frac {n-k-1}{2(1-n)}},
\end{align*}
where the function $U=U(z)$ satisfies the second-order pantograph-type ODE:
\begin{align*}
&\frac 1{1-n}U+\frac {n-k-1}{2(1-n)}zU'_z=a(U^kU'_z)'_z+U^nf(q^{\frac{1}{1-n}}W/U),\\
&W=U(sz),\quad \ s=pq^{\ts\frac {n-k-1}{2(1-n)}}.
\end{align*}

\medskip
\textit{\textbf{Equation 9.}}  The equation containing an arbitrary function
\begin{align}
u_t=a(e^{\lambda u}u_x)_x+e^{\mu u}f(u-w),\quad \ w=u(px,qt),
\label{eq:abc}
\end{align}
for $\mu\not=0$ admits the exact solution
\begin{align*}
u(x,t)=U(z)-\frac 1\mu \ln t,\quad \ z=xt^{\ts\frac{\lambda-\mu}{2\mu}},
\end{align*}
where the function $U=U(z)$ is described by the nonlinear pantograph-type ODE:
\begin{align*}
&\frac{\lambda-\mu}{2\mu}zU'_z-\frac 1\mu=a(e^{\lambda U}U'_z)'_z+e^{\mu U}f\Bl(U-W+\frac 1\mu\ln q\Br)=0,\\
&W=U(sz),\quad \ s=pq^{\ts\frac{\lambda-\mu}{2\mu}}.
\end{align*}

\hypertarget{hlink-ss:2.4}{}
\subsection{Nonlinear Equations Containing Two Arbitrary Functions}\label{ss:2.4}

\textit{\textbf{Equation 10.}}
The equation with varying transfer coefficient of general form
\begin{align}
u_t=a[uf'_u(u)u_x]_x+\frac 1{f'_u(u)}[bf(u)+cf(w)+d],\quad w=u(px,qt),
\label{eq:016**}
\end{align}
admits the functional separable solution in implicit form
\begin{align}
f(u)=\varphi(t)x+\psi(t),
\label{eq:016***}
\end{align}
where the functions $\varphi=\varphi(t)$ and $\psi=\psi(t)$ satisfy the pantograph-type ODEs:
\begin{align*}
\varphi'_t&=b\varphi+cp\bar\varphi,\quad \ \bar\varphi=\varphi(qt),\\
\psi'_t&=b\psi+c\bar\psi+d+a\varphi^2,\quad \bar\psi=\psi(qt).
\end{align*}

\begin{Remark}
 More general than Eq. \eqref{eq:016**}, the equation in which the pantograph term $w=u(px,qt)$ is replaced by $w=u(p_1x+p_0,q_1t+q_0)$
also has the functional separable solution \eqref{eq:016***}.
The more complex functional differential equation in which
$w=u(p_1x+p_0,\eta(t))$, where $\eta(t)$ \arbf, also has a solution of the same form.
\end{Remark}

\medskip
\textit{\textbf{Equation 11.}}  For $q=p$, a nonlinear pantograph-type
reaction-diffusion equation of the general form
\begin{align*}
u_{t}=[g(u)u_x]_x+F(u,w),\quad \ w=u(px,pt),
\end{align*}
admits the traveling-wave solution
\begin{align*}
u(x,t)=U(z), \quad \ z=kx-\lambda t,
\end{align*}
where $k$ and $\lambda$ \arbs\ and the function $U=U(z)$ is described by the nonlinear pantograph-type ODE:
$$
k^2[g(U)U'_z]'_z+\lambda U'_z+F(U,W)=0,\quad \ W=U(pz).
$$

\hypertarget{hlink-s:3}{}
\section{Solutions of Pantograph-Type PDEs That Contain Unknown Functions $u=u(x,t)$ and $w=u(x,qt)$}\label{s:3}

\textit{Preliminary remarks.}
This section does not include equations and their solutions that can be obtained by setting $p=1$ in the equations and solutions considered in Section~\ref{s:2}.

\hypertarget{hlink-ss:3.1}{}
\subsection{Equations Containing Free Parameters}\label{ss:3.1}

\textit{\textbf{Equation 12.}}  The equation with logarithmic nonlinearity
\begin{align}
u_t=au_{xx}+u(b\ln u+c\ln w+d),\quad \ w=u(x,qt),
\label{eq:004}
\end{align}
admits the functional separable solution
\begin{align*}
u(x,t)=\exp[\psi_2(t)x^2+\psi_1(t)x+\psi_0(t)],
\end{align*}
where the functions $\psi_n=\psi_n(t)$ are described by the nonlinear system of pantograph-type ODEs:
\begin{align*}
&\psi_2^\prime=4a\psi_2^2+b\psi_2+c\bar\psi_2,\quad \bar\psi_2=\psi_2(qt),\\
&\psi_1^\prime=4a\psi_1\psi_2+b\psi_1+c\bar\psi_1,\quad \bar\psi_1=\psi_1(qt),\\
&\psi_0^\prime=a[\psi_1^2+2\psi_2]+b\psi_0+c\bar\psi_0+d,\quad \bar\psi_0=\psi_0(qt).
\end{align*}

\medskip
\textit{\textbf{Equation 13.}}  The equation with logarithmic nonlinearity
\begin{align}
u_t=au_{xx}+u(b\ln^2 u+c\ln u+d\ln w+s),\quad \ w=u(x,qt),
\label{eq:005}
\end{align}
depending on the sign of the product $ab$, admits two functional separable solutions given below.

$1^\circ$. The solution for $ab>0$:
\begin{align*}
&u(x,t)=\exp[\psi_1(t)\varphi(x)+\psi_2(t)],\\
&\varphi(x)=A \cos(\lambda x)+B\sin(\lambda x),\quad \lambda=\sqrt{b/a},
\end{align*}
where $A$ and $B$ \arbs\ and the functions $\psi_n=\psi_n(t)$ are described by the nonlinear system of pantograph-type ODEs:
\begin{align*}
\psi_1^\prime&=2b\psi_1\psi_2+(c-b)\psi_1+d\bar\psi_1,\quad \bar\psi_1=\psi_1(qt),\\
\psi_2^\prime&=b(A^2+B^2)\psi_1^2+b\psi_2^2+c\psi_2+d\bar\psi_2+s,\quad \bar\psi_2=\psi_2(qt).
\end{align*}

$2^\circ$. The solution for $ab<0$:
\begin{align*}
&u(x,t)=\exp[\psi_1(t)\varphi(x)+\psi_2(t)],\\
&\varphi(x)=A \cosh(\lambda x)+B\sinh(\lambda x),\quad \lambda=\sqrt{-b/a},
\end{align*}
where $A$ and $B$ \arbs\ and the functions $\psi_n=\psi_n(t)$ are described by the nonlinear system of pantograph-type ODEs:
\begin{align*}
\psi_1^\prime&=2b\psi_1\psi_2+(c-b)\psi_1+d\bar\psi_1,\quad \bar\psi_1=\psi_1(qt),\\
\psi_2^\prime&=b(A^2-B^2)\psi_1^2+b\psi_2^2+c\psi_2+d\bar\psi_2+s,\quad \bar\psi_2=\psi_2(qt).
\end{align*}
For $A=\pm B$, we have $\varphi(x)=Ae^{\pm\lambda x}$. In this case, the second equation of the system becomes independent and the first one becomes linear in $\psi_1$.

\begin{Remark}
Eqs. \eqref{eq:004} and \eqref{eq:005} and their solutions can be substantially generalized
if the pantograph term $w=u(x,qt)$ is replaced by
$w=u(x,\eta(t))$, where $\eta(t)$ \arbf \ (the solutions are sought in the same form, see \cite{pol2013}).
\end{Remark}

\hypertarget{hlink-ss:3.2}{}
\subsection{Equations Linear in Derivatives and Containing One Arbitrary Function}\label{ss:3.2}

\textit{\textbf{Equation 14.}} The equation containing an arbitrary function
\begin{align*}
u_t=au_{xx}+f(u-w),\quad \ w=u(x,qt),
\end{align*}
admits the additive separable solution
\begin{align*}
u(x,t)=C_1x^2+C_2x+\psi(t),
\end{align*}
where $C_1$ and $C_2$ are arbitrary real constants and the function $\psi=\psi(t)$ is described by the nonlinear first-order pantograph-type ODE:
\begin{align*}
\psi'_t=2aC_1+f(\psi-\bar\psi),\quad \bar\psi=\psi(qt).
\end{align*}

\medskip
\textit{\textbf{Equation 15.}} The equation containing an arbitrary function
\begin{align}
u_t=au_{xx}+bu+ f(u-w),\quad \ w=u(x,q t),
\label{eq:006}
\end{align}
depending on the sign of the product  $ab$, admits two additive separable solutions given below.

$1^\circ$. The solution for $ab<0$:
\begin{align*}
u(x,t)=A\cosh(\lambda x)+B\sinh(\lambda x)+\psi(t),\quad \lambda = \sqrt{-b/a},
\end{align*}
where $A$ and $B$ \arbs\ and the function $\psi=\psi(t)$ is described by the nonlinear first-order pantograph-type ODE:
\begin{align}
\psi'_t=b\psi+f(\psi-\bar\psi),\quad \bar\psi=\psi(q t).
\label{eq:006:1}
\end{align}

$2^\circ$. The solution for $ab>0$:
\begin{align*}
u(x,t)=A\cos(\lambda x)+B\sin(\lambda x)+\psi(t),\quad \lambda = \sqrt{b/a},
\end{align*}
where $A$ and $B$ \arbs\ and the function $\psi=\psi(t)$ is described by the nonlinear first-order pantograph-type ODE \eqref{eq:006:1}.

Note that Eq. \eqref{eq:006} and its solutions can be substantially generalized if the pantograph term $w=u(x,qt)$
is replaced by  $w=u(x,\eta(t))$, where $\eta(t)$ \arbf \ \ (the solutions are sought in the same form, see \cite{pol2013,pol2014c}).

\medskip
\textit{\textbf{Equation 16.}} The equation containing an arbitrary function
\begin{align*}
u_t=au_{xx}+uf(w/u),\quad \ w=u(x,qt),
\end{align*}
admits several multiplicative separable solutions given below.

$1^\circ$. The solution
\begin{align*}
u(x,t)=[A\cosh(\lambda x)+B\sinh(\lambda x)]\psi(t),
\end{align*}
where $A$, $B$, and $\lambda$ \arbs\
and the function $\psi=\psi(t)$ is described by the nonlinear first-order pantograph-type ODE:
\begin{align*}
\psi'_t=a\lambda^2\psi+\psi f(\bar\psi/\psi),\quad \bar\psi=\psi(qt).
\end{align*}

$2^\circ$. The solution
\begin{align*}
u(x,t)=[A\cos(\lambda x)+B\sin(\lambda x)]\psi(t),
\end{align*}
where $A$, $B$, and $\lambda$ \arbs\
and the function $\psi=\psi(t)$ is described by the nonlinear first-order pantograph-type ODE:
\begin{align*}
\psi'_t=-a\lambda^2\psi+\psi f(\bar\psi/\psi),\quad \bar\psi=\psi(qt).
\end{align*}

$3^\circ$. The degenerate solution
\begin{align*}
u(x,t)=(Ax+B)\psi(t),
\end{align*}
where $A$, $B$, and $\lambda$ \arbs\
and the function $\psi=\psi(t)$ is described by the nonlinear first-order pantograph-type ODE:
\begin{align*}
\psi'_t=\psi f(\bar\psi/\psi),\quad \bar\psi=\psi(qt).
\end{align*}

\medskip
\textit{\textbf{Equation 17.}} The equation containing an arbitrary function
\begin{align}
u_t=au_{xx}+bu \ln u+u f(w/u),\quad \ w=u(x,qt),
\label{eq:007*}
\end{align}
admits the multiplicative separable solution
\begin{align*}
u(x,t)=\varphi(x)\psi(t).
\end{align*}
where the functions $\varphi=\varphi(x)$ and $\psi=\psi(t)$ are described by the nonlinear second-order ODE and the nonlinear first-order pantograph-type ODE:
\begin{equation}
\begin{aligned}
&a\varphi''_{xx}=C_1\varphi-b\varphi\ln\varphi,\\
&\psi^\prime_t=C_1\psi+\psi f(\bar\psi/\psi)+b\psi\ln\psi,\quad \bar\psi=\psi(qt);
\end{aligned}
\label{eq:008}
\end{equation}
$C_1$ \arb.

The first equation of Eq.~\eqref{eq:008} is autonomous; its general solution can be obtained in implicit form. A particular one-parameter solution of this equation has the form
$$
\varphi=\exp\biggl[-\frac{b}{4a}(x+C_2)^2+\frac{C_1}{b}+\frac12\biggr],
$$
where $C_2$ \arb.

Note that Eq.\eqref{eq:007*} and its solution can be substantially generalized if the pantograph term $w=u(x,qt)$ is replaced by $w=u(x,\eta(t))$, where $\eta(t)$ \arbf \
\ (the solution is sought in the same form, see \cite{pol2013,pol2014c}).

\hypertarget{hlink-ss:3.3}{}
\subsection{More Complex Nonlinear Equations Containing One Arbitrary Function}\label{ss:3.3}

\textit{\textbf{Equation 18.}} The equation with varying transfer coefficient of power-law type
\begin{align}
u_t=a(u^ku_x)_x+uf(w/u), \quad w=u(x,qt)
\label{eq:009}
\end{align}
admits the multiplicative separable solution
$$
u(x,t)=\varphi(x)\psi(t),
$$
where the functions $\varphi=\varphi(x)$ and $\psi=\psi(t)$ are determined from the ODE and the pantograph-type ODE:
\begin{align*}
&a(\varphi^k\varphi'_x)'_x=b\varphi,\\
&\psi'_t=b\psi^{k+1}+\psi f\bl(\bar\psi/\psi\br),\quad \ \bar\psi=\psi(qt);
\end{align*}
$b$ \arb.

\medskip
\textit{\textbf{Equation 19.}} The equation with varying transfer coefficient of power-law type
\begin{align}
u_t=a(u^ku_x)_x+uf(w/u)+bu^{k+1},\quad w=u(x,qt),
\label{eq:010}
\end{align}
admits three multiplicative separable solutions given below.

$1^\circ$. The solution for $b(k+1)>0$:
\begin{equation*}
u(x,t)=[C_1\cos(\beta x)+C_2\sin(\beta x)]^{1/(k+1)}\psi(t),\quad \ \beta=\sqrt{b(k+1)/a},
\end{equation*}
where $C_1$ and $C_2$ \arbs\ and the function $\psi=\psi(t)$ is described by the pantograph-type ODE:
\begin{equation}
\psi'_t=\psi f\bl(\bar\psi/\psi\br),\quad \ \bar\psi=\psi(qt).
\label{ode:1001}
\end{equation}

$2^\circ$. The solution for $b(k+1)<0$:
\begin{equation*}
u(x,t)=[C_1\exp(-\beta x)+C_2\exp(\beta x)]^{1/(k+1)}\psi(t),\quad \ \beta=\sqrt{-b(k+1)/a},
\end{equation*}
where $C_1$ and $C_2$ \arbs\ and the function $\psi=\psi(t)$ is described by the pantograph-type ODE~\eqref{ode:1001}.

$3^\circ$. The solution for $k=-1$:
\begin{equation*}
u(x,t)=C_1\exp\Bl(-\frac b{2a}x^2+C_2x\Br)\psi(t),
\end{equation*}
where $C_1$ and $C_2$ \arbs\ and the function $\psi=\psi(t)$ is described by the pantograph-type ODE~\eqref{ode:1001}.

\begin{Remark}
Eqs. \eqref{eq:009} and \eqref{eq:010} and their solutions can be substantially generalized
if the pantograph term $w=u(x,qt)$ is replaced by $w=u(x,\eta(t))$, where $\eta(t)$ \arbf
\ (the solutions are sought in the same form, see \cite{pol2014f}).
\end{Remark}

\medskip
\textit{\textbf{Equation 20.}}  The equation with varying transfer coefficient of power-law type
\begin{align*}
u_{t}=a(u^ku_x)_x+u^{k+1}f(w/u),\quad \ w=u(x,qt),
\end{align*}
admits the exact solution
\begin{align*}
u(x,t)=t^{-1/k}\varphi(z),\quad \ z=x+\lambda\ln t,
\end{align*}
where $\lambda$ \arb\ and the function $\varphi=\varphi(z)$ satisfies the second-order ODE with constant delay
\begin{align*}
a(\varphi^k\varphi'_z)'_z-\lambda\varphi'_z+\frac 1k\varphi+\varphi^{k+1}f(q^{-1/k}\bar\varphi/\varphi)=0,\quad \
\bar\varphi=\varphi(z+\lambda\ln q).
\end{align*}

\medskip
\textit{\textbf{Equation 21.}} The equation with varying transfer coefficient of power-law type
\begin{align}
u_t=a(u^ku_x)_x+b+u^{-k}f(u^{k+1}-w^{k+1}),\quad w=u(x,qt),
\label{eq:011}
\end{align}
admits the functional separable solution
\begin{equation*}
u(x,t)=\Bl[\psi(t)-\frac{b(k+1)}{2a}x^2+C_1x+C_2\Br]^{1/(k+1)},
\end{equation*}
where $C_1$ and $C_2$ \arbs\ and the function $\psi=\psi(t)$ is described by the pantograph-type ODE:
\begin{equation*}
\psi'_t=(k+1)f(\psi-\bar\psi\br),\quad \ \bar\psi=\psi(qt).
\end{equation*}

\medskip
\textit{\textbf{Equation 22.}}
The equation with varying transfer coefficient of exponential type
\begin{align}
u_t=a(e^{\lambda u}u_x)_x+f(u-w),\quad w=u(x,qt),
\label{eq:013a}
\end{align}
admits the additive separable solution
\begin{equation*}
u(x,t)=\frac 1\lambda\ln(Ax^2+Bx+C)+\psi(t),
\end{equation*}
where $A$, $B$, and $C$ \arbs\ and the function $\psi(t)$ is described by the pantograph-type ODE:
\begin{equation*}
\psi'=2a(A/\lambda)e^{\lambda \psi}+f(\psi-\bar\psi), \quad \ \bar\psi=\psi(qt).
\end{equation*}

\medskip
\textit{\textbf{Equation 23.}}
The equation with varying transfer coefficient of exponential type
\begin{align}
u_t=a(e^{\lambda u}u_x)_x+f(u-w)+be^{\lambda u},\quad w=u(x,qt),
\label{eq:013}
\end{align}
admits two additive separable solutions.

$1^\circ$. The solution for $b\lambda>0$:
\begin{equation*}
u(x,t)=\frac 1\lambda \ln[C_1\cos(\beta x)+C_2\sin(\beta x)]+\psi(t),\quad \ \beta=\sqrt{b\lambda/a},
\end{equation*}
where $C_1$ and $C_2$ \arbs\ and the function $\psi=\psi(t)$ is described by the pantograph-type ODE:
\begin{equation}
\psi'_t=f\bl(\psi-\bar\psi\br),\quad \ \bar\psi=\psi(qt).
\label{ode:1002}
\end{equation}

$2^\circ$. The solution for $b\lambda<0$:
\begin{equation*}
u(x,t)=\frac 1\lambda \ln[C_1\exp(-\beta x)+C_2\exp(\beta x)]+\psi(t),\quad \ \beta=\sqrt{-b\lambda/a},
\end{equation*}
where $C_1$ and $C_2$ \arbs\ and the function $\psi=\psi(t)$ is described by the pantograph-type ODE \eqref{ode:1002}.

\begin{Remark}
Eqs. \eqref{eq:011}--\eqref{eq:013} and their solutions can be substantially generalized
if the pantograph term $w=u(x,qt)$ is replaced by
$w=u(x,\eta(t))$, where $\eta(t)$ \arbf \ (the solutions are sought in the same form, see \cite{pol2014f}).
\end{Remark}

\medskip
\textit{\textbf{Equation 24.}}
The equation with varying transfer coefficient of exponential type
\begin{align}
u_t=a(e^{\lambda u}u_x)_x+b+e^{-\lambda u}f(e^{\lambda u}-e^{\lambda w}),\quad w=u(x,qt),
\label{eq:014}
\end{align}
admits the functional separable solution
\begin{equation*}
u(x,t)=\frac 1\lambda\ln\Bl[\psi(t)-\frac{b\lambda}{2a}x^2+C_1x+C_2\Br],
\end{equation*}
where $C_1$ and $C_2$ \arbs\ and the function $\psi=\psi(t)$ is described by the pantograph-type ODE:
\begin{equation*}
\psi'_t=\lambda f\bl(\psi-\bar\psi\br),\quad \ \bar\psi=\psi(qt).
\end{equation*}

\medskip
\textit{\textbf{Equation 25.}}
The equation with varying transfer coefficient of logarithmic type
\begin{align}
u_t=[(a\ln u+b)u_x]_x-cu\ln u+uf(w/u),\quad w=u(x,qt),
\label{eq:015}
\end{align}
admits two multiplicative separable solutions
$$
u(x,t)=\exp(\pm\sqrt{c/a}\,x)\psi(t),
$$
where the function $\psi=\psi(t)$ is described by the pantograph-type ODE:
\begin{equation*}
\psi'_t=c(1+b/a)\psi+\psi f\bl(\bar\psi/\psi\br),\quad \ \bar\psi=\psi(qt).
\end{equation*}

\hypertarget{hlink-ss:3.4}{}
\subsection{Nonlinear Equations Containing Two Arbitrary Functions}\label{ss:3.4}

\textit{\textbf{Equation 26.}}
The equation with varying transfer coefficient of general form
\begin{align}
u_t=a[g'_u(u)u_x]_x+b+\frac 1{g'_u(u)}f\bl(g(u)-g(w)\br),\quad w=u(x,qt),
\label{eq:017}
\end{align}
admits the functional separable solution in implicit form
\begin{equation*}
g(u)=\psi(t)-\frac{b}{2a}x^2+C_1x+C_2,
\end{equation*}
where the function $\psi=\psi(t)$ is described by the pantograph-type ODE  \eqref{ode:1002}.

\medskip
\textit{\textbf{Equation 27.}}
The equation with varying transfer coefficient of general form
\begin{align}
u_t=a[g'_u(u)u_x]_x+bg(u)+\frac{g(u)}{g'_u(u)}f\bl(g(w)/g(u)\br),\quad w=u(x,qt),
\label{eq:018}
\end{align}
admits two functional separable solutions in implicit form.

$1^\circ$. The solution for $ab>0$:
\begin{equation*}
g(u)=\bl[C_1\cos(\lambda x)+C_2\sin(\lambda x)\br]\psi(t),\quad \ \lambda=\sqrt{b/a},
\end{equation*}
where $C_1$ and $C_2$ \arbs\ and  the function $\psi=\psi(t)$ is described by the pantograph-type ODE  \eqref{ode:1001}.

$2^\circ$. The solution for  $ab<0$:
\begin{equation*}
g(u)=\bl[C_1\exp(-\lambda x)+C_2\exp(\lambda x)\br]\psi(t),\quad \ \lambda=\sqrt{-b/a},
\end{equation*}
where $C_1$ and $C_2$ \arbs\ and  the function $\psi=\psi(t)$ is described by the pantograph-type ODE \eqref{ode:1001}.

\begin{Remark}
Eqs.  \eqref{eq:014}--\eqref{eq:018} and their solutions can be substantially generalized
if the pantograph term $w=u(x,qt)$ is replaced by $w=u(x,\eta(t))$, where $\eta(t)$ \arbf \ \ (the solutions are sought in the same form, see \cite{pol2014f}).
\end{Remark}

\hypertarget{hlink-s:4}{}
\section{Solutions of Pantograph-Type PDEs That Contain Unknown Functions $u=u(x,t)$ and $w=u(px,t)$}

\textit{Preliminary remarks.}
This section does not include equations and their solutions that can be obtained by setting $q=1$ in the equations and solutions considered in Section~\ref{s:2}.

\hypertarget{hlink-ss:4.1}{}
\subsection{Equations Linear in Derivatives}

\textit{\textbf{Equation 28.}} The equation containing an arbitrary function
\begin{align}
u_t=au_{xx}+f(u-w),\quad \ w=u(px,t),
\label{eq:0001}
\end{align}
admits the additive separable solution
\begin{align*}
u(x,t)=Ct+\varphi(x),
\end{align*}
where $C$ \arb\ and the function $\varphi=\varphi(x)$ is described by the nonlinear second-order pantograph-type ODE:
\begin{align*}
a\varphi''_{xx}-C+f(\varphi-\bar\varphi)=0,\quad \bar\varphi=\varphi(px).
\end{align*}

\medskip
\textit{\textbf{Equation 29.}} The equation containing an arbitrary function
\begin{align}
u_t=au_{xx}+bu+ f(u-w),\quad \ w=u(px,t),
\label{eq:0002}
\end{align}
admits the additive separable solution
\begin{align*}
u(x,t)=Ce^{bt}+\varphi(x),
\end{align*}
where $C$ \arb\ and the function $\varphi=\varphi(x)$ is described by the nonlinear second-order pantograph-type ODE:
\begin{align*}
a\varphi''_{xx}+b\varphi+f(\varphi-\bar\varphi)=0,\quad \ \bar\varphi=\varphi(px).
\end{align*}

\medskip
\textit{\textbf{Equation 30.}} The equation containing an arbitrary function
\begin{align}
u_t=au_{xx}+uf(w/u),\quad \ w=u(px,t),
\label{eq:0003}
\end{align}
admits the multiplicative separable solution
\begin{align*}
u(x,t)=e^{\lambda t}\varphi(x),
\end{align*}
where $\lambda$ \arb\ and the function $\varphi=\varphi(x)$ is described by the nonlinear second-order pantograph-type ODE:
\begin{align*}
a\varphi''_{xx}+\varphi[f(\bar\varphi/\varphi)-\lambda]=0,\quad \ \bar\varphi=\varphi(px).
\end{align*}

\medskip
\textit{\textbf{Equation 31.}} The equation containing an arbitrary function
\begin{align}
u_t=au_{xx}+bu \ln u+u f(w/u),\quad \ w=u(px,t),
\label{eq:0004}
\end{align}
admits the multiplicative separable solution
\begin{align*}
u(x,t)=\exp(Ce^{bt})\varphi(x),
\end{align*}
where $C$ \arb\ and the function $\varphi=\varphi(x)$ is described by the nonlinear pantograph-type ODE:
\begin{align*}
a\varphi''_{xx}+b\varphi\ln\varphi+\varphi f(\bar\varphi/\varphi)=0,\quad \ \bar\varphi=\varphi(px).
\end{align*}

\begin{Remark}
 Eqs. \eqref{eq:0001}--\eqref{eq:0004} and their solutions can be substantially generalized
if the pantograph term $w=u(px,t)$ is replaced by  $w=u(\xi(x),t)$, where $\xi(x)$ \arbf \ (the solutions are sought in the same form).
\end{Remark}

\hypertarget{hlink-ss:4.2}{}
\subsection{More Complex Nonlinear Equations}

\textit{\textbf{Equation 32.}} The equation with varying transfer coefficient of power-law type
\begin{align}
u_t=a(u^ku_x)_x+uf(w/u), \quad w=u(px,t)
\label{eq:xz}
\end{align}
admits the exact solution
$$
u(x,t)=e^{2\lambda t}U(z),\quad \ z=e^{-k\lambda t}x,
$$
where $\lambda$ \arb\ and  the function $U=U(z)$ is determined from the pantograph-type ODE:
\begin{align*}
2\lambda U-k\lambda zU'_z=a(U^kU'_z)'_z+Uf(W/U),\quad \ W=U(pz).
\end{align*}

\medskip
\textit{\textbf{Equation 33.}} The equation with varying transfer coefficient of power-law type
\begin{align}
u_t=a(u^ku_x)_x+u^{k+1}f(w/u),\quad \ w=u(px,t).
\label{eq:xz*}
\end{align}
admits the multiplicative separable solution
$$
u(x,t)=t^{-1/k}\varphi(x),
$$
where the function $\varphi=\varphi(x)$ is described by the pantograph-type ODE:
$$
a(\varphi^k\varphi'_x)'_x+\frac 1k\varphi+\varphi^{k+1}f(\bar\varphi/\varphi)=0,\quad \ \bar\varphi=\varphi(px).
$$


\medskip
\textit{\textbf{Equation 34.}} The equation with varying transfer coefficient of power-law type
\begin{align}
u_t=a(u^ku_x)_x+bu^{-k}+f(u^{k+1}-w^{k+1}),\quad \ w=u(px,t),
\label{eq:xz***}
\end{align}
admits the functional separable solution
$$
u(x,t)=\bl[b(k+1)t+\varphi(x)\br]^{\ts\frac 1{k+1}},
$$
where the function $\varphi=\varphi(x)$ is described by the pantograph-type ODE:
$$
a\varphi''_{xx}+(k+1)f(\varphi-\bar\varphi)=0,\quad \ \bar\varphi=\varphi(px).
$$


\medskip
\textit{\textbf{Equation 35.}} The equation with varying transfer coefficient of exponential type
\begin{align}
u_t=a(e^{\lambda u}u_x)_x+e^{\lambda u}f(u-w),\quad \ w=u(px,t),
\label{eq:xz**}
\end{align}
admits the additive separable solution
$$
u(x,t)=-\frac 1\lambda\ln t+\varphi(x),
$$
where the function $\varphi=\varphi(x)$ is described by the pantograph-type ODE:
$$
a(e^{\lambda \varphi}\varphi'_x)'_x+\frac 1\lambda+e^{\lambda\varphi}f(\varphi-\bar\varphi)=0,\quad \ \bar\varphi=\varphi(px).
$$


\medskip
\textit{\textbf{Equation 36.}} The equation with varying transfer coefficient of general form
\begin{align}
u_t=[f'_u(u)u_x]_x+\frac a{f'_u(u)}+g(f(u)-f(w)),\quad \ w=u(px,t),
\label{eq:700}
\end{align}
admits the generalized traveling wave solution in implicit form
$$
f(u)=at+\varphi(x),
$$
where the function $\varphi=\varphi(x)$ is described by the pantograph-type ODE:
$$
\varphi''_{xx}+g(\varphi-\bar\varphi)=0,\quad \ \bar\varphi=\varphi(px).
$$

\begin{Remark}
 Eqs. \eqref{eq:xz*}--\eqref{eq:700} and their solutions can be substantially generalized
if the pantograph term $w=u(px,t)$ is replaced by  $w=u(\xi(x),t)$, where $\xi(x)$ \arbf \ (the solutions are sought in the same form).
\end{Remark}

\hypertarget{hlink-s:5}{}
\section{Some Generalizations}

\hypertarget{hlink-ss:5.1}{}
\subsection{Nonlinear Multi-Pantograph Type PDEs}

The equations and their solutions considered above can be generalized to the case of nonlinear multi-pantograph type PDEs of the  form
\begin{align*}
u_{t}=[g(u)u_x]_x+F(u,w_1,\dots,w_m),\quad \ w_j=u(p_jx,q_jt), \ \ j=1,\,\dots,\,m.
\end{align*}
Two such equations with exact solutions are given below.

\medskip
\textit{\textbf{Equation 37.}}
Consider the nonlinear multi-pantograph type PDE:
\begin{align}
u_{t}=a(u^ku_x)_x+u^nf(w_1/u,\dots,w_m/u),\quad \ w_j=u(p_jx,q_jt),
\label{eq-z*}
\end{align}
which generalizes Eq.\eqref{eq-z}.
For $n\not=1$, Eq. \eqref{eq-z*} admits a self-similar solution of the same form as Eq.~\eqref{eq-z}:
\begin{align*}
u(x,t)=t^{\ts\frac 1{1-n}}U(z),\quad \ z=xt^{\ts\frac {n-k-1}{2(1-n)}},
\end{align*}
where the function $U=U(z)$ satisfies the second-order multi-pantograph type ODE:
\begin{align*}
&\frac 1{1-n}U+\frac {n-k-1}{2(1-n)}zU'_z=a(U^kU'_z)'_z+U^nf\bl(q_1^{\ts\frac 1{1-n}}W_1/U,\dots,q_m^{\ts\frac 1{1-n}}W_m/U\br),\\
&W_j=U(s_jz),\quad \ s_j=p_jq_j^{\ts\frac {n-k-1}{2(1-n)}}, \ \ j=1,\,\dots,\,m.
\end{align*}

\medskip
\textit{\textbf{Equation 38.}}
The nonlinear multi-pantograph type PDE, more general than Eq.~\eqref{eq:xz},
\begin{align*}
u_t=a(u^ku_x)_x+uf(w_1/u,\dots,w_m/u), \quad w_j=u(p_jx,t),\quad \ w_j=u(p_jx,t),
\end{align*}
admits a solution of the same form as Eq.~\eqref{eq:xz}:
$$
u(x,t)=e^{2\lambda t}U(z),\quad \ z=e^{-k\lambda t}x,
$$
where $\lambda$ \arb\ and the function $U=U(z)$ is determined from the multi-pantograph type ODE:
\begin{align*}
2\lambda U-k\lambda zU'_z=a(U^kU'_z)'_z+Uf(W_1/U,\dots,W_m/U),\quad \ W_j=U(p_jz).
\end{align*}

\hypertarget{hlink-ss:5.2}{}
\subsection{Nonlinear Pantograph-Type PDEs Containing Unknown Functions with Dilation or Contraction of Arguments in the Derivative}

One can also consider more general than Eq.~\eqref{eq:007} nonlinear pantograph-type reaction-diffusion equations of the form
\begin{align*}
u_{t}=[G(u,w)u_x]_x+F(u,w).
\end{align*}
Below are three examples of exact solutions to such equations.

\medskip
\textit{\textbf{Equation 39.}}
Consider the pantograph-type diffusion equation with varying transfer coefficient
\begin{align}
u_t=[G(u,w)u_x]_x,\quad \ w=u(px,qt).
\label{eq:yy}
\end{align}

$1^\circ$. In the general case, this equation admits the self-similar solution
\begin{align*}
u(x,t)=U(z), \quad \ z=xt^{-1/2},
\end{align*}
where the function $U=U(z)$ is described by the nonlinear pantograph-type ODE:
\begin{align*}
[G(U,W)U'_z]'_z+\tfrac12zU'_z=0,\quad \ W=U(sz),\quad s=pq^{-1/2}.
\end{align*}

$2^\circ$. For $q=p$, Eq. \eqref{eq:yy} admits the traveling-wave solution
$$
u(x,y)=U(z),\quad \ z=kx-\lambda t,
$$
where $k$ and $\lambda$ \arbs\ and the function $U=U(z)$ is described by the first-order pantograph-type ODE:
$$
k^2G(U,W)U'_z+\lambda U=C,\quad \ W=U(pz);
$$
$C$ \arb.

\medskip
\textit{\textbf{Equation 40.}}
Consider now the equation
\begin{align}
u_t=a(e^{\lambda w}u_x)_x+e^{\mu u}f(u-w),\quad \ w=u(px,qt),
\label{eqqq}
\end{align}
which is obtained from Eq.~\eqref{eq:abc} by replacing the function $e^{\lambda u}$ with $e^{\lambda w}$.
For $\mu\not=0$, this equation, like Eq.~\eqref{eq:abc},
admits an exact solution of the form
\begin{align}
u(x,t)=U(z)-\frac 1\mu \ln t,\quad \ z=xt^{\ts\frac{\lambda-\mu}{2\mu}},
\label{eqqq*}
\end{align}
where the function $U=U(z)$ is described by the nonlinear pantograph-type ODE:
\begin{align*}
&\frac{\lambda-\mu}{2\mu}zU'_z-\frac 1\mu=aq^{-\lambda/\mu}(e^{\lambda W}U'_z)'_z+e^{\mu U}f\Bl(U-W+\frac 1\mu\ln q\Br),\\
&W=U(sz),\quad \ s=pq^{\ts\frac{\lambda-\mu}{2\mu}}.
\end{align*}

\medskip
\textit{\textbf{Equation 41.}}
The equation, more general than Eq.~\eqref{eqqq},
$$
u_t=[e^{\lambda u}g(u-w)u_x]_x+e^{\mu u}f(u-w),\quad \ w=u(px,qt),
$$
depending on two arbitrary functions $f(y)$ and $g(y)$ also has a solution of the form~\eqref{eqqq*}.

\hypertarget{hlink-ss:5.3}{}
\subsection{Principle of Analogy of Solutions}\label{ss:5.3}

A fairly general method for constructing exact solutions of partial functional differential pantograph-type equations is based on the use of the following principle.

\textbf{The principle of analogy of solutions}.
\textit{The structure of exact solutions of functional differential equations of the form
\begin{equation}
\begin{aligned}
&F(x,t,u,u_x,u_t,u_{xx},u_{xt},u_{tt},\dots,w,w_x,w_t,w_{xx},w_{xt},w_{tt},\dots)=0,\\
&w=u(px,qt),
\end{aligned}
\label{eq:xyzz}
\end{equation}
is often (but not always) determined by the structure of solutions of simpler PDEs:
\begin{align}
F(x,t,u,u_x,u_t,u_{xx},u_{xt},u_{tt},\dots,u,u_x,u_t,u_{xx},u_{xt},u_{tt},\dots)=0.
\label{eq:xyzzz}
\end{align}}

Equation \eqref{eq:xyzzz} does not contain the unknown function with dilated or contracted arguments;
it is obtained from \eqref{eq:xyzz} by formally replacing $w$ with $u$.

Most of the solutions to nonlinear pantograph-type reaction-diffusion equations described in this article have been obtained using the principle of analogy.
Let us illustrate what has been said with specific examples.

\textbf{Example 1.}
Consider the pantograph-type PDE with
power-law nonlinearities
\begin{align}
u_t=au_{xx}+bu^mw^k,\quad \ w=u(px,qt).
\label{bbb5}
\end{align}

Following the principle of analogy of solutions, we set $w=u$ in
Equation~\eqref{bbb5}. As a result, we arrive at the simpler PDE
\begin{align}
u_t=au_{xx}+bu^{m+k}.
\label{bbb6}
\end{align}
This equation admits a self-similar solution of the form \cite{dor1982}:
\begin{align*}
u(x,t)=t^{\ts\frac 1{1-m-k}}U(z), \quad \ z=xt^{-1/2},\quad \ k\not=1-m.
\end{align*}
Using the principle of analogy, we seek the solution of the pantograph-type PDE~\eqref{bbb5} also in the form \eqref{bbb6}.
As a result, for the function $U=U(z)$ we get the nonlinear
pantograph-type ODE:
\begin{align*}
&aU''_{zz}+\frac12zU'_z-\frac 1{1-m-k}U+bq^{\frac k{1-m-k}}U^mW^k=0,\\
&W=U(sz),\quad \ s=pq^{-1/2}.
\end{align*}
\medskip

\textbf{Example 2.}
Let us now consider the pantograph-type PDE
equation with exponential nonlinearities
\begin{align}
u_t=au_{xx}+be^{\mu u+\lambda w},\quad \ w=u(px,qt).
\label{bbb7}
\end{align}
Following the principle of
analogy of solutions, we set $w=u$ in Equation~\eqref{bbb7}. As a result,
we arrive at the simpler PDE
\begin{align}
u_t=au_{xx}+be^{(\mu+\lambda)u}.
\label{bbb8}
\end{align}
This equation admits an invariant solution of the form \cite{dor1982}:
\begin{align*}
u(x,t)=U(z)-\frac 1{\mu+\lambda} \ln t,\quad \ z=xt^{-1/2},\quad \ \mu\not=-\lambda.
\end{align*}
Using the principle of analogy, we seek the solution of the pantograph-type PDE~\eqref{bbb7} in the form \eqref{bbb8}.
As a result, for the function $U=U(z)$ we obtain the nonlinear
pantograph-type ODE:
\begin{align*}
&aU''_{zz}+\frac 12zU'_z+\frac 1{\mu+\lambda}+
bq^{\ts-\frac{\lambda}{\mu+\lambda}} e^{\mu U+\lambda W}=0,\\
&W=U(sz),\quad \ s=pq^{-1/2}.
\end{align*}

Note that the principle of analogy of solutions can also be successfully applied to construct exact solutions of functional differential wave-type PDEs \cite{polsor2020p} and PDEs of higher
orders that contain unknown functions with dilated or contracted arguments.

\hypertarget{hlink-s:6}{}
\section{Brief Conclusions}\label{s:6}

For the first time, we have described
exact solutions of various classes of nonlinear pantograph-type reaction-diffusion PDEs that, in addition to the unknown function
$u=u(x,t)$, also contain functions with dilated or contracted arguments of the form $u(px,t)$, $u(x,qt)$, and $u(px,qt)$,
where $p$ and $q$ are the free scaling parameters.
We have presented examples of nonlinear pantograph-type PDEs that admit traveling-wave solutions, self-similar solutions, additive and
multiplicative separable solutions, functional separable solutions, and some other exact solutions.
Special attention is paid to nonlinear pantograph-type PDEs of a rather general form, which contain one or two arbitrary functions.
A number of exact solutions of more complex nonlinear functional differential equations with varying delay, which arbitrarily depends
on time or spatial coordinate, are described. We have formulated the principle of analogy of solutions
that allows effectively constructing exact solutions of nonlinear pantograph-type PDEs.
The presented exact solutions can be used as test problems to evaluate the accuracy of numerical and approximate analytical
methods for solving the corresponding nonlinear initial-boundary value problems for PDEs with varying delay of pantograph type.

\hypertarget{hlink-s:7}{}
\renewcommand{\refname}{References}\label{s:7}

\end{document}